\documentclass[twocolumn]{aastex62}
\usepackage{graphicx}
\usepackage{subfigure}
\usepackage{color}
\usepackage{amsmath}
\usepackage{lineno}

\begin{document}

\title{Cosmic Evolution of Stellar-mass Black Hole Merger Rate in Active Galactic Nuclei}
\author{Y. Yang}
\affiliation{Department of Physics, University of Florida, PO Box 118440, Gainesville, FL 32611-8440, USA}
\author{I. Bartos}
\thanks{imrebartos@ufl.edu}
\affiliation{Department of Physics, University of Florida, PO Box 118440, Gainesville, FL 32611-8440, USA}
\author{Z. Haiman}
\affiliation{Department of Astronomy, Columbia University in the City of New York, 550 W 120th St., New York, NY 10027, USA}
\author{B. Kocsis}
\affiliation{E\"otv\"os University, Institute of Physics, P\'azm\'any P. s. 1/A, Budapest, 1117, Hungary}
\author{S. M\'arka}
\affiliation{Department of Physics, Columbia University in the City of New York, 550 W 120th St., New York, NY 10027, USA}
\author{H. Tagawa}
\affiliation{Institute of Physics, E{\"o}tv{\"o}s University, P{\'a}zm{\'a}ny P.s., Budapest, 1117, Hungary}

\begin{abstract}
Binary black hole mergers encode information about their environment and the astrophysical processes that led to their formation. Measuring the redshift dependence of their merger rate will help probe the formation and evolution of galaxies and the evolution of the star formation rate. Here we compute the cosmic evolution of the merger rate for stellar-mass binaries in the disks of Active Galactic Nuclei (AGNs). We focus on recent evolution out to redshift $z=2$, covering the accessible range of current Earth-based gravitational-wave observatories. On this scale, the AGN population density is the main contributor to redshift-dependence. We find that the AGN-assisted merger rate does not meaningfully evolve with redshift, differentiating this channel from field binaries and some other dynamical formation scenarios. 
\end{abstract}

\section{Introduction}

Observations by the LIGO and Virgo gravitational-wave detectors \citep{aLIGO2015,Acernese_2014} show a high rate of stellar-mass black hole mergers of $\sim 10-100$\,Gpc$^{-3}$yr$^{-1}$ \citep{PhysRevX.9.031040}. Despite the rapidly growing number of detections, however, the origin of these black hole mergers is currently not known. Possible formation mechanisms include isolated stellar binary evolution \citep{2012ApJ...759...52D,2014MNRAS.442.2963K,2018MNRAS.474.2959G,bavera2019origin} and chance encounters in dense stellar clusters such as galactic nuclei or globular clusters \citep{2000ApJ...528L..17P,2006ApJ...637..937O,2015PhRvL.115e1101R,2016ApJ...824L..12O,Samsing_2014,Zhang_2019}. 

Active galactic nuclei (AGN) represent a unique environment in which the interaction of a dense cusp of stellar-mass black holes near the galactic center \citep{2018Natur.556...70H} and the dense AGN accretion disk result in dramatically altered merger rates and properties \citep{2014MNRAS.441..900M,2017ApJ...835..165B,2017MNRAS.464..946S}. As black holes orbiting the central, supermassive black hole cross the AGN disk, they experience friction that can align their orbit with the disk \citep{2017ApJ...835..165B}. Additional black holes can be born within the AGN disk due to gravitational fragmentation \citep{2017MNRAS.464..946S}. Once in the disk, black holes migrate inward where their density further increases, enabling a higher rate of interaction \citep{2019arXiv190910517T}. 

AGN disks act as black hole assembly lines that collect and concentrate black holes in small volumes, enhancing their merger rate. The resulting rate can be a significant fraction of the total merger rate observed by LIGO and Virgo with estimates ranging within $\sim10^{-3}-10^{2}$\,Gpc$^{-3}$yr$^{-1}$ \citep{2019ApJ...876..122Y,2019arXiv190910517T,2018ApJ...866...66M}.

The properties of black holes within AGN-assisted mergers are expected to be different from other formation channels, enabling observational probes. Heavier black holes will be overrepresented in mergers within AGN disks compared to the black hole initial mass function as they can more efficiently align their orbit with the disk \citep{2019ApJ...876..122Y}. In addition, as multiple black holes are driven towards the same small volume within the disk, consecutive mergers of the same black hole, or so-called hierarchical mergers, are common \citep{PhysRevLett.123.181101}. Such hierarchical mergers will result in characteristic high black hole spins, which will typically be aligned or anti-aligned with the binary orbit \citep{PhysRevLett.123.181101,2019arXiv190910517T}. Two  particular black hole mergers recorded by LIGO-Virgo so far, GW170729 \citep{PhysRevX.9.031040} and GW170817A \citep{2019arXiv191009528Z}, has the characteristically high mass and aligned spin expected from hierarchical mergers in AGN disks, albeit they are also consistent with other hierarchical formation channels \citep{PhysRevLett.123.181101,2020ApJ...890L..20G}. 

Beyond the properties of the black holes themselves, AGNs provide other means to probe this population. Mergers in AGN disks are only expected in galaxies with active nuclei, which can be used to statistically differentiate them from other formation channels \citep{2017NatCo...8..831B}. Additionally, merger in a gas-rich environment may produce detectable electromagnetic emission \citep{2017ApJ...835..165B,2019ApJ...884L..12Y,2019ApJ...884L..50M}.

Here we investigated a distinct property of a binary merger population: its rate evolution with redshift. Redshift dependence can be used to differentiate between different formation channels and to better understand the underlying mechanisms that result in binary formation and merger. While the binary merger rate's redshift dependence has been previously explored for different binary formation scenarios (e.g., \citealt{2018ApJ...863L..41F}), our analysis is the first such investigation for the AGN channel.

The paper is organized as follows. In \S\ref{sec:models} we examine the expected redshift evolution for different formation channels. In \S\ref{sec:GW} we discuss the conversion of rate densities to detection rates. In \S\ref{sec:results} we present our results. We conclude in \S\ref{sec:conclusion}.

\section{Binary formation channels} \label{sec:models}

In this section we compute the expected cosmic evolution of the black hole merger rate density for the AGN channel, and review the expected evolution for field binaries and globular clusters from the literature. 

\subsection{The AGN channel}

The black hole merger rate in AGNs is proportional to the AGN population density $n_{\rm AGN}(z)$, where $z$ is redshift. It can be evaluated through the AGN luminosity function (LF) $\phi_{\rm L}(L,z)$, which is a function of redshift and AGN luminosity $L$. The bolometric AGN LF can be fitted as \citep{2020arXiv200102696S}:
\begin{equation}
    \phi_{\rm L}(L,z)=\frac{\phi_{*}(z)}{[\frac{L}{L_{*}(z)}]^{\gamma_{\rm 1}(z)}+[\frac{L}{L_{*}(z)}]^{\gamma_{\rm 2}(z)}}{\rm Mpc^{-3}}
\end{equation}
where
\begin{align}
&\gamma_1(z)=a_0T_0(1+z)+a_1T_1(1+z)+a_2T_2(1+z)\\
&\gamma_2(z)=\frac{2b_0}{(\frac{1+z}{3})^{b_1}+(\frac{1+z}{3})^{b_2}}\\
&\log{\L_{*}(z)}=\frac{2c_0}{(\frac{1+z}{3})^{c_1}+(\frac{1+z}{3})^{c_2}}\\
&\log{\phi_{*}(z)}=d_0T_0(1+z)+d_1T_{1}(1+z)
\end{align}
where $T_n$ is the n-th order Chebyshev polynomial. The best fit for the 11 parameters in this model are: $\{a_0,a_1,a_2;b_0,b_1,b_2;c_0,c_1,c_2;d_0,d_1\}$=\{0.8396, -0.2519, 0.0198; 2.5432, -1.0528, 1.1284; 13.0124, -0.5777, 0.4545; -3.5148, -0.4045\}. \\
The direct integration of $\phi_{\rm L}(L,z)$ will yield the AGN density $n_{\rm AGN}(z)$. However, the lower end of the LF is subject to large uncertainty, thus we introduced a cutoff $L_{\rm min}$ when integrating. On the other hand, the mass of SMBHs can be correlated with the AGN luminosity via:
\begin{equation}
    \frac{M_{\bullet}}{M_{\odot}}=3.17\times10^{-5}\frac{1-\epsilon}{\dot{m}}\frac{L}{L_{\odot}} \label{M-L relation}
\end{equation}
where $\epsilon$ is the radiation efficiency of the SMBH, $\dot{m}=\dot{M_{\bullet}}/\dot{M}_{\rm Edd}$, $\dot{M_{\bullet}}$ is the accretion rate of the SMBH, and $\dot{M}_{\rm Edd}=L_{\rm Edd}/\epsilon c^2$ is the Eddington rate. 

Eq. \ref{M-L relation} could be rewritten to give a relation between the normalized accretion rate $\dot{m}$ and Eddington ratio $\lambda=L/L_{\rm Edd}$:
\begin{equation}
    \dot{m}=(1-\epsilon)\lambda
\end{equation}
The Eddington ratio $\lambda$ is found to take to form \citep{2017A&A...600A..64T}:
\begin{equation}
    P(\lambda|L,z)=f_{\rm uno}P_1(\lambda|z)+f_{\rm obs}P_2(\lambda|z)
\end{equation}
where $f_{\rm uno}=1-f_{\rm obs}$ is the fraction of unobscured (type-1) AGN and $f_{\rm obs}$ is the fraction of obscured (type-2) AGN. $P_1$ and $P_2$ are the Eddington ratio distributions of type-1 and type-2 AGNs, respectively.\\
$P_1(\lambda|z)$ follows a log-normal distribution:
\begin{equation}
    P_1(\lambda|z)=\frac{1}{2\pi\sigma(z)\lambda}e^{-[\ln{\lambda}-\ln{\lambda_{\rm c}(z)}]^2/2\sigma^2(z)}
\end{equation}
with $\log{\lambda_{\rm c}(z)}=\max(-1.9+0.45z,\log(0.03))$ and $\sigma(z)=\max(1.03-0.15z,0.6)$.\\
$P_2(\lambda|z)$ follows a gamma distribution with a cut-off at low-Eddington luminosities:
\begin{equation}
    P_2(\lambda|z)=N_2(z)\lambda^{\alpha(z)}e^{-\lambda/\lambda_0}
\end{equation}
where $\lambda_0=1.5$ and $N_2(z)$ is the normalization factor. The slope of the power law part, $\alpha(z)$, takes the form:
\begin{align}
   \alpha(z)=\left\{ \begin{array}{rll}
&\mbox{$-0.6$}  & \mbox{ $z < 0.6$} \\
&\mbox{$-0.6/(0.4+z)$}  & \mbox{ $z \geq 0.6$}\\
\end{array} \right.
\end{align}
We assume that the Eddington ratio distribution has a cut-off ($\lambda_{\rm l}$) at low-Eddington luminosities, which is fixed to be $10^{-4}$ in our study. Our results below are not sensitive to this choice, which we confirmed for the $\lambda_{\rm l}=10^{-4}-10^{-2}$ range.

The fraction $f_{\rm obs}$ can be parameterized as a function of X-ray luminosity $L_{\rm X}$ and redshift \citep{2014ApJ...786..104U}:
\begin{equation}
    f_{\rm obs}=\frac{(1+f_{\rm CTK})\psi(L_{\rm X},z)}{1+f_{\rm CTK}\psi(L_{\rm X},z)}
\end{equation}
where $f_{\rm CTK}$ is the relative number density of compton thick (CTK, $\log{N_{\rm H}}>24$) AGNs to that of compton thin (CTN, $\log{N_{\rm H}}=20-24$) AGNs, $N_{\rm H}$ is the neutral hydrogen column density in unit of ${\rm cm}^{-2}$. We assume $f_{\rm CTK}=1$ is this work. $\psi$ is the fraction of obscured AGNs($\log{N_{\rm H}}=20-22$) in total CTN AGNs and can be expressed as:
\begin{align}
   \psi(L_{\rm X},z)=&\min(\psi_{\rm max},\max(\psi_{43.75}(z)\nonumber\\
   &-\beta(\log{L_{\rm X}}-43.75),\psi_{\rm min}))
\end{align}
where we adopt $\psi_{\rm max}=0.84$, $\psi_{\rm min}=0.2$ and $\beta=0.24$. $\psi_{43.75}(z)$ can be written as:
\begin{align}
   \psi_{43.75}(z)=\left\{ \begin{array}{rll}
&\mbox{$0.43(1+z)^{0.48}$}  & \mbox{ $z < 2$} \\
&\mbox{$0.43(1+2)^{0.48}$}  & \mbox{ $z \geq 2$}\\
\end{array} \right.
\end{align}
Since $f_{\rm obs}$ is dependent on X-ray luminosity, we need to convert the bolometric luminosity to the X-ray luminosity using  a  bolometric  correction \citep{2004MNRAS.351..169M}:
\begin{equation}
    \log{(L/L_{\rm X})}=1.54+0.24\xi + 0.012\xi^2-0.0015\xi^3 \label{bolo-correction}
\end{equation}
with $\xi=\log{L/L_{\odot}}-12$. 
lower limit of the Eddington ratio. The distribution function of $\lambda$ is independent of the redshift, therefore the average Eddington ratio is a constant for $z \lesssim 1$, as shown by other studies \citep{2017MNRAS.471.1976G}.

We additionally need the BH merger rate ($\Gamma$) of a single AGN. \cite{2019ApJ...876..122Y} showed that several factors affect the BH merger rate, of which $\dot{m}$ is the dominant one. We assumed that the mean number($N_{\rm disk}$) of stellar black holes in AGN disk is a univariate function of $\dot{m}$ and obtained a power law fit ($M_{\bullet}=10^6M_{\odot},\epsilon=0.1$):
\begin{equation}
    N_{\rm disk}(\dot{m})=5.5\,\dot{m}^{1/3}
\end{equation}
We assume that the black holes in AGN disks will hierarchically merge in the migration traps \citep{PhysRevLett.123.181101} and the number of stellar black holes in AGN disks follows a Poisson distribution with mean value of $N_{\rm disk}$. Consequently, the average BH merger rate is:
\begin{equation}
    \Gamma(\dot{m})=(N_{\rm disk}(\dot{m})-1+e^{-N_{\rm disk}(\dot{m})})/\tau_{\rm AGN}
\end{equation}
where $\tau_{\rm AGN}=10^7{\rm yr}$ is the AGN lifetime.\\
We note that this dependence is different for several subdominant merger processes associated with AGNs ( see \citealt{2019arXiv190910517T} for a comparison of different contributing processes), however we found that our results below only weakly depend on the specific form of $\Gamma(\dot{m})$, therefore other AGN-related processes will not alter our results below. 

Combining these above factors, we arrived at an expression for the redshift-dependence of the cosmic BH merger rate in AGN:
\begin{equation}
    R_{\rm AGN}(z)=\int\limits_{L\in I_{\rm L}}\phi_{\rm L}d\log{L
    }\int\limits_{\lambda_{\rm l} }^{1}\Gamma(\dot{m})P(\lambda|L,z)d\lambda\,.
\end{equation}
Here, the integral domain of the luminosity is $I_{\rm L}=[L_{\rm min},3.15\times10^{14}L_{\odot}]$.

\begin{figure}
   \centering  
   \includegraphics[width=0.47\textwidth]{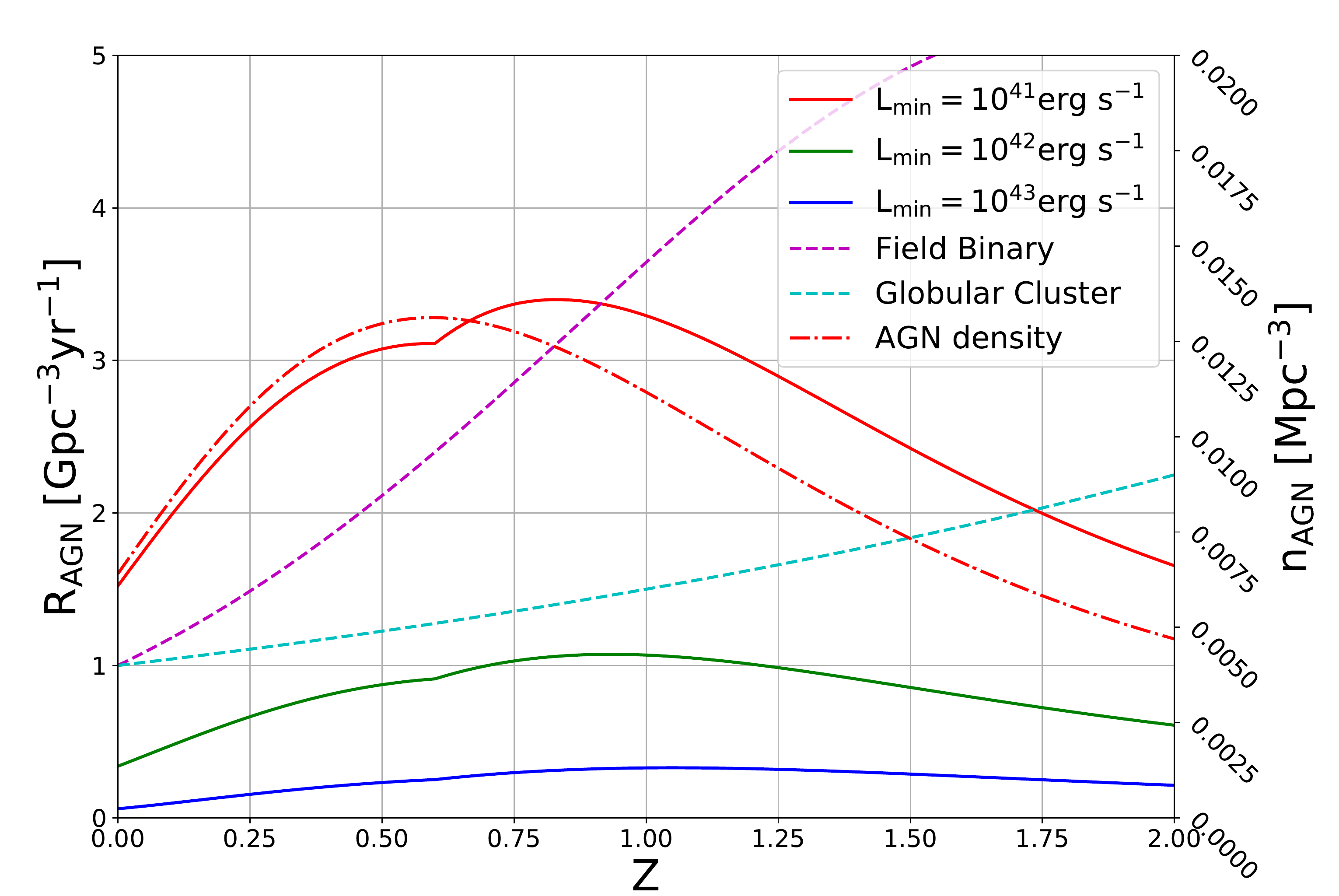}
   \caption{The cosmic AGN-assisted black hole merger rate as a function of redshift for several choices of $L_{\rm min}$. We adopted a radiation efficiency of $\epsilon=0.1$. We compared their merger rate with the merger rate (dashed lines) of BBHs from globular clusters and field binaries, whose local merger rate at $z=0$ is fixed to be 1Gpc$^{-3}$yr$^{-1}$. We also added the AGN density distribution (dotted dashed line) with $L_{\rm min}=10^{41}$erg s$^{-1}$.  }   
\label{fig:cosmic merger rate}
\end{figure}

We show $R_{\rm AGN}(z)$ in Fig. \ref{fig:cosmic merger rate} for multiple choices of $L_{\rm min}$. We see that the distribution only weakly depends on $z$, with a maximum around $z=0.8$ which is about a factor of two greater than the minimum at $z=0$. We also see that the choice of $\L_{\rm min}$ does not meaningfully affect the normalized redshift distribution of $R_{\rm AGN}(z)$, although it does change the magnitude of merger rate density. Therefore, in the following, we adopted $L_{\rm min}=10^{41}{\rm erg\ s^{-1}}$ as our fiducial model.

\subsection{Field binaries}

We also compared our model with other formation channels. The first considered mechanism is the field binary channel. We assumed that the formation rate density of field binaries in comoving volume follows the low-metallicity star formation rate (SFR; \citealt{2018ApJ...863L..41F}):
\begin{align}
    \rho_{\rm FB}(z)&\propto\psi_{\rm MD}(z)f_{\rm Z}(z)\nonumber\\
    &\propto\frac{(1+z)^{2.7}}{1+(\frac{1+z}{2.9})^{5.6}}\gamma{(0.84,\left(\frac{Z}{Z_\odot}\right)^210^{0.3z})}
\end{align}
where $\psi_{\rm MD}(z)$ is the Madau-Dickinson SFR \citep{2014ARA&A..52..415M} and $f_{\rm Z}(z)$ is the fraction of star formation occurring at metallicity smaller than $Z$, $\gamma(s,x)$ is the lower incomplete gamma function. 

Since the binaries do not merge immediately after their formation, we adopted a time-delay model to estimate the merger rate density:
\begin{equation}
R_{\rm FB}=\int_{z_{\rm m}}^{\infty}\rho_{\rm FB}(z_{\rm f})p(t(z_{\rm f})-t(z_{\rm m}))\frac{dt}{dz_{\rm f}}dz_{\rm f}
\end{equation}
where $p(t)$ is the distribution of time delay and $t(z)$ is the cosmological look back time. We assumed that the time delay had a flat distribution in log space between $50$\,Myr and $15$\,Gyr. The mass distribution was assumed to be \citep{2018ApJ...863L..41F}:
\begin{equation}
f_{\rm FB}(m_{\rm 1},m_{\rm 2})=\frac{1-\alpha}{M_{\rm max}^{1-\alpha}-(5M_{\odot})^{1-\alpha}}\frac{m_{\rm 1}^{-\alpha}}{m_{\rm 1}-5M_{\odot}}
\end{equation}
where $5M_{\odot}<m_{\rm 2}<m_{\rm 1}<M_{\rm max}$.


\subsection{Globular clusters}

Another channel we considered is the dynamical formation in globular clusters (e.g. \citealt{2018PhRvL.121p1103F}). We found that the merger rate density at $z\lesssim 2$ can be parametrized as $R_{\rm GC}=18.6\times(\frac{3}{2})^z$Gpc$^{-3}$yr$^{-1}$. We also considered other merger rate density distribution(e.g. \citealt{2016PhRvD..93h4029R,2018ApJ...866L...5R}), but it does not change at $z\lesssim 2$ by more than a factor 3. Thus, we adopted the exponential parameterization in our analysis. The mass function was assumed to be the same as above.

\section{Conversion to detection rate} \label{sec:GW}

In order to compare the expected redshift evolutions to black hole merger observations via gravitational waves, we need to convert the expected merger rate distribution $R_{\rm merger}$ to detection rate distribution $R_{\rm det}$ using the sensitive distance range of LIGO-Virgo. 
\begin{equation}
    R_{\rm det}(z)=\frac{R_{\rm merger}}{1+z}\frac{dV_{\rm c}}{dz}\int P_{\rm det}(\mathcal{M})f(m_{\rm 1},m_{\rm 2})dm_{\rm 1}dm_{\rm 2}. \label{detection rate}
\end{equation}
Here, $V_{\rm c}$ is the co-moving volume and $f(m_{\rm 1},m_{\rm 2})$ is the mass function of binary black holes in AGN disks. 
$P_{\rm det}(\mathcal{M})$ is the probability of an event at redshift $z$ with detector-frame chirp mass 
\begin{equation}
\mathcal{M} =(1+z)\frac{(m_{\rm 1}m_{\rm 2})^{3/5}}{(m_{\rm 1}+m_{\rm 2})^{1/5}}
\end{equation}
being detected by advanced LIGO. In our calculations, we assumed that an event is detectable when its signal-to-noise ratio (SNR) is greater than 8. The SNR of a black hole merger is  \citep{2010ApJ...716..615O}:
\begin{equation}
    \rho=8\omega\frac{370\,\mbox{Mpc}}{D_{\rm L}(z)}\left(\frac{\mathcal{M}}{M_{\odot}}\right)^{5/6}
\end{equation}
where $D_{\rm L}(z)$ is the luminosity distance corresponding to redshift $z$, and $\omega \in [0,1]$ is a geometrical factor determined by the position and orientation of the binary system\footnote{The tabular data of $P(\omega)$ can be found online at  \href{http://www.phy.olemiss.edu/~berti/research.html}{http://www.phy.olemiss.edu/~berti/research.html}}. We adopt the noise power spectral density of LIGO at its design sensitivity \citep{2016PhRvD..93k2004M,2011PhRvD..84h4037A} in the evaluation of the SNR above. 

\cite{2019ApJ...876..122Y} found that the AGN disk can significantly change the initial mass function (IMF) of merging black holes. The hierarchical black hole mergers will then alter the mass distribution of binary black holes \citep{PhysRevLett.123.181101} since the mass of one component (the remnant of the previous merger) of the binary will increase as the hierarchical merging process continues. We adopted the weighted average binary mass distribution in their work($N_{\rm disk}=2.5$):
\begin{equation}
    \bar{f}(m_{\rm 1},m_{\rm 2})=\sum_{\rm n=1}^{\infty}P_{\rm n}f_{\rm n}(m_{\rm 1},m_{\rm 2})
\end{equation}
where $P_{\rm n}$ and $f_{\rm n}$ are the fraction and mass distribution of n-th generation, respectively.


\section{Results} \label{sec:results}

We calculated the expected redshift distributions of detected events for the models described in Section \ref{sec:models} using the conversion described in Section \ref{sec:GW}. We then compared these distributions to the reconstructed cosmic evolution for black hole mergers observed through gravitational waves by LIGO-Virgo during the O1 and O2 observing periods (Model B in \citealt{LIGOScientific:2018jsj}). 

\begin{figure}
   \centering
   \subfigure{\includegraphics[width=0.5\textwidth]{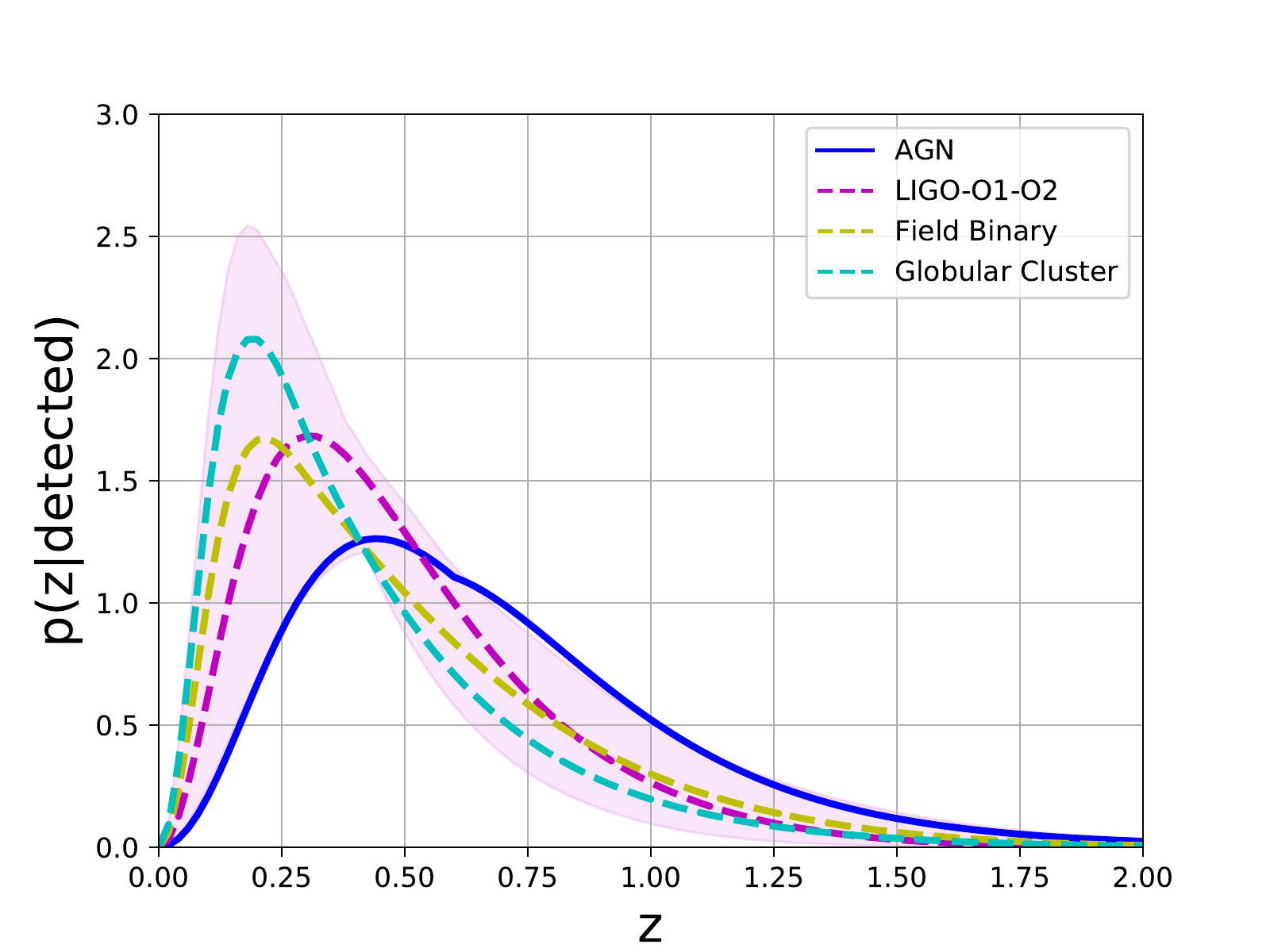}}\hfill
\caption{The expected redshift distribution of detected gravitational wave events by Advanced LIGO-Virgo at design sensitivity. The fitted parameters of the model for LIGO's detections (LIGO-O1-O2) are $m_{\rm min}=7.8M_{\odot}, m_{\rm max}=40.8M_{\odot}, \alpha=1.3, \beta_{\rm q}=6.9$, and the shaded region is the $90\%$ credible intervals \citep{LIGOScientific:2018jsj}. For the field binary channel, we assume that the formation rate density of field binaries follows the low-metallicity star formation rate and adopt a time-delay model to evaluate the merger rate density. For the dynamical mergers in globular cluster, we postulate the merger rate density $R_{\rm GC}\propto(3/2)^{z}$. }   
\label{fig:redshiftdistribution}
\end{figure}

Our results are shown in Fig. \ref{fig:redshiftdistribution}. We see that the expected rate evolution for LIGO-Virgo observations is currently uncertain and is essentially consistent with all three formation channel models considered here. Looking at the LIGO-Virgo distribution using its expected value, we see that the observed distribution peaks at a higher redshift than the field-binary and globular cluster channels, but at a lower redshift than our AGN model. Taking this expected distribution at face value, the observed distribution is consistent with having a 40\% AGN and 60\% field-binary contribution.

\section{Conclusion} \label{sec:conclusion}

We computed the expected redshift distribution of the merger rate of stellar-mass black hole mergers in AGN disks. We found that the distribution is close to being uniform out to $z\approx1$, which is distinct from our expectations for field binaries and for some other dynamical merger scenarios, such as in globular clusters. This distinct evolution, together with other differences, can help differentiate between the possible origins of binary mergers and help probe their environment.

\begin{acknowledgements}
The authors are thankful to the University of Florida and Columbia University in
the City of New York for their generous support. The Columbia Experimental Gravity group is grateful for the generous support of the National Science Foundation under grant PHY-1708028. This project was supported by funds from the European Research
Council (ERC) under the European Union’s Horizon 2020 research and innovation programme under grant agreement No 638435 (GalNUC) and by the Hungarian National Research, Development, and Innovation Office grant NKFIH KH-125675 (to BK). ZH acknowledges support from NASA grant NNX15AB19G and NSF grant 1715661.
\end{acknowledgements}

\bibliography{Refs}
\end{document}